\documentclass[a4paper,12pt]{article}

\usepackage{ifpdf}

\newif\ifpdf
\ifx\pdfoutput\undefined
  \pdffalse
\else
  \pdfoutput=1
  \pdftrue
\fi

\RequirePackage{xspace} %
\RequirePackage{subfigure} %
\RequirePackage[centertags]{amsmath} %
\RequirePackage{amssymb}
\RequirePackage{wrapfig} %
\RequirePackage{calc} %
\RequirePackage{ifthen}
\RequirePackage{tabularx} %
\RequirePackage{flafter} %
\RequirePackage{fancyhdr} %

\ifpdf
  \RequirePackage[pdftex]{color}%
  \RequirePackage{colortbl}%
  \RequirePackage{array}%
  \RequirePackage[pdftex]{graphicx}

  \RequirePackage[ pdftex, plainpages = false, pdfpagelabels,
                 pdfpagelayout = useoutlines,
                 bookmarks,
                 breaklinks = true,
                 linktocpage,
                 pagebackref,                      
                 colorlinks = true,
                 linkcolor = blue,
                 urlcolor  = blue,
                 citecolor = blue,
                 anchorcolor = blue,
                 hyperindex = true,
                 hyperfigures
                 ]{hyperref}

\else
  \RequirePackage{color}
  \RequirePackage{colortbl}
   \RequirePackage{array}
  \RequirePackage[dvips]{graphicx}
  \RequirePackage{hyperref}
  \usepackage{rotating}
\fi

\usepackage{makeidx} 
\usepackage{setspace} 
\usepackage{rotating} 
\usepackage{ecltree}
\usepackage{epic}
\usepackage{supertabular}  
\usepackage{color}
\usepackage{exscale}
\usepackage{fontenc}
\usepackage{ifthen}
\usepackage{latexsym}
\usepackage{makeidx}
\usepackage{syntonly}
\usepackage{inputenc}
\usepackage{graphicx}
\usepackage{setspace}
\usepackage{caption2}
\usepackage[english]{babel}
\usepackage[square, comma,numbers,sort&compress]{natbib}
\usepackage{hypernat}
\usepackage{boxedminipage}
\usepackage{framed}
\usepackage{longtable}
\usepackage[all]{hypcap}    
\usepackage{algorithm2e}
\usepackage{algorithmic}
\usepackage{lscape}
\usepackage{pdflscape}
\usepackage{mathtools}

\newcommand{\note}[1]{\marginpar[left]{\singlespace \tiny #1}}

\renewcommand{\sectionmark}[1]%
      {\markright{\thesection\ #1}} 

\renewcommand{\note}[1]{}

\newcommand{\CIF}     {\centering \includegraphics[width=2.7in]} %
\newcommand{\Vmin}    {\vspace{-0.2cm}} %
\newcommand{\Hs}      {\hspace{-0.5cm}} %

\doublespace

\begin{document}
\begin{center}
{\Large Variational approach for the flow of Ree-Eyring and Casson fluids in pipes}
\par\end{center}{\Large \par}

\begin{center}
Taha Sochi
\par\end{center}

\begin{center}
{\scriptsize University College London, Department of Physics \& Astronomy, Gower Street, London,
WC1E 6BT \\ Email: t.sochi@ucl.ac.uk.}
\par\end{center}

\begin{abstract}
\noindent The flow of Ree-Eyring and Casson non-Newtonian fluids is investigated using a
variational principle to optimize the total stress. The variationally-obtained solutions are
compared to the analytical solutions derived from the Weissenberg-Rabinowitsch-Mooney equation and
the results are found to be identical within acceptable numerical errors and modeling
approximations.

\vspace{0.3cm}

\noindent Keywords: fluid mechanics; tube flow; Ree-Eyring; Casson;
Weissenberg-Rabinowitsch-Mooney; variational method; total stress optimization.

\par\end{abstract}

\section{Introduction} \label{Introduction}

Many analytical and numerical methods have been invented and used, especially in the recent
decades, for modeling, simulating and analyzing the flow of various types of fluid under diverse
flow conditions in conduits of different shapes and geometries and with different mechanical
properties. These methods include the application of first principles of mechanics, the use of
Navier-Stokes equation, the application of lubrication approximation, the use of
Weissenberg-Rabinowitsch-Mooney relation, stochastic methods, and finite element and spectral
methods \cite{Miekisz1963, ShahLBook1978, PapanastasiouGABook1999, LetelierSC2002, Livescu2012,
GeorgiouK2013, SochiElastic2013, SochiElasticTubeNew2014, ShaikhM2014}.

Recently, we proposed \cite{SochiVariational2013} the use of a new method which is based on
applying the Euler-Lagrange variational principle to optimize the total stress of fluid in the flow
conduit for the purpose of obtaining relations for the flow of generalized Newtonian fluids.
Although the method is general, it lends itself more easily to the flow in rigid straight uniform
conduits with circularly-shaped cross sections as the flow in these cases is essentially
unidirectional.

There, the method was demonstrated by using a straight cylindrical tube geometry with six fluid
rheological models: Newtonian, power law, Bingham plastic, Herschel-Bulkley, Carreau and Cross. The
proposed method can be applied analytically, when the variational equation can be solved in all its
stages by analytical means, as well as numerically in part when some stages are difficult or
impossible to solve analytically and hence standard numerical methods, like numerical integration
by Simpson's rule and solving implicit equations by bisection methods, can be used. It was also
demonstrated that although the variational principle in its proposed form is restricted to fluids,
and hence it does not include viscoplastic materials which behave like solids prior to their yield
point, the method is a good approximation when applied to the flow of viscoplastic materials of low
yield stress value since the effect of the non-yielded plug region on the total flow is minimal.

In a later study \cite{SochiSlitPaper2014} we provided more support to the variational method by
successful application to the flow in a plane long thin slit geometry using two rheological models:
Newtonian and power law where the well-known analytical solutions for these fluids were obtained
from the application of the variational principle.

In the present study, we further investigate and validate the variational method through its
application to the flow of two non-Newtonian fluid models, Ree-Eyring and Casson, in rigid straight
uniform pipes with circular cross sectional shape. The first of these is a pseudoplastic model
which is normally used to describe the flow of shear thinning fluids, while the second is a
viscoplastic model used to describe the rheology of yield stress materials. The method is applied
analytically in part and numerically in other parts. The investigation related to the Casson model
provides further support to the fact that the variational method is a good approximation for the
viscoplastic materials characterized by a low value of yield tress.

\section{Variational Method}

Before outlining the variational method and its application to the flow of Ree-Eyring and Casson
fluids, we state the assumptions and approximations that have been considered in the model
development. In this investigation, we assume a laminar, incompressible, steady, pressure-driven,
fully-developed flow of time-independent, purely-viscous fluids that can be described by the
generalized Newtonian fluid model which is given by

\begin{equation}\label{GNF}
    \tau = \mu \gamma
\end{equation}
where $\tau$ is the shear stress, $\mu$ is the generalized Newtonian viscosity, and $\gamma$ is the
rate of shear strain. As for the conduit, we assume a rigid straight tube with a uniform cross
section (i.e. constant shape and area along the axial dimension) of circular shape.

The variational method for obtaining the flow relations for generalized Newtonian fluids is based
on the minimization of the total stress in the flow conduit using the Euler-Lagrange variational
principle to obtain the rate of shear strain. This is then followed by obtaining the flow velocity
profile by integrating the rate of strain with respect to the tube radius, which is followed by
integrating the flow velocity profile with respect to the cross sectional area to obtain the
volumetric flow rate. Although the first step in this method is based on analytical manipulation of
the derived variational principle, the subsequent stages can be manipulated either analytically or
numerically. This allows more flexibility in the application of the variational method as will be
vividly demonstrated in the forthcoming paragraphs and sections.

From the basic principles and definitions of the fluid mechanics of generalized Newtonian fluids,
the total stress arising from the flow profile of a fluid flowing in a tube of radius $R$ is given
by

\begin{equation}
\tau_{t}=\int_{\tau_{m}}^{\tau_{w}}d\tau=\int_{0}^{R}\frac{d\tau}{dr}dr=\int_{0}^{R}\frac{d}{dr}\left(\mu\gamma\right)dr=\int_{0}^{R}\left(\gamma\frac{d\mu}{dr}+\mu\frac{d\gamma}{dr}\right)dr\label{totalStress}
\end{equation}
where $\tau_{t}$ is the total shear stress, $\tau_{m}$ and $\tau_{w}$ are the shear stress at the
tube center and wall respectively, and $r$ is the radius at an arbitrary point on the tube cross
section.

The total stress, as given by Equation \ref{totalStress}, can be optimized by applying the
Euler-Lagrange variational principle which, in one of its diverse forms, is given by

\begin{equation}
\frac{d}{dx}\left(f-y'\frac{\partial f}{\partial y'}\right)-\frac{\partial f}{\partial x}=0
\end{equation}
where the symbols in the last equation correspond to the symbols of the stated flow problem as
follow
\begin{equation}
x\equiv r,\,\,\,\,\, y\equiv\gamma,\,\,\,\,\,
f\equiv\gamma\frac{d\mu}{dr}+\mu\frac{d\gamma}{dr},\,\,\,\mathrm{\,\, and}\,\,\,\,\,\frac{\partial
f}{\partial
y'}\equiv\frac{\partial}{\partial\gamma'}\left(\gamma\frac{d\mu}{dr}+\mu\frac{d\gamma}{dr}\right)=\mu
\end{equation}
that is

\begin{equation}\label{prev}
\frac{d}{dr}\left(\gamma\frac{d\mu}{dr}+\mu\frac{d\gamma}{dr}-\mu\frac{d\gamma}{dr}\right)-\frac{\partial}{\partial
r}\left(\gamma\frac{d\mu}{dr}+\mu\frac{d\gamma}{dr}\right)=0
\end{equation}
Considering the fact that for the considered flow systems

\begin{equation}
\gamma\frac{d\mu}{dr}+\mu\frac{d\gamma}{dr}=G
\end{equation}
where $G$ is a constant, it can be shown that Equation \ref{prev} can be reduced to two independent
variational forms

\begin{equation}\label{MainVar1}
\frac{d}{d r}\left(\gamma\frac{d\mu}{dr}\right)=0
\end{equation}
and

\begin{equation}\label{MainVar}
\frac{d}{d r}\left(\mu\frac{d\gamma}{dr}\right)=0
\end{equation}

In the following two subsections we use the second form where we outline the application of the
variational method, as summarized in Equation \ref{MainVar}, to the Ree-Eyring and Casson fluids.
We then validate the variationally-obtained flow solutions by comparing them to the
analytically-derived solutions from the Weissenberg-Rabinowitsch-Mooney (WRM) equation.

\subsection{Ree-Eyring Fluids}

For the Ree-Eyring fluids, the rheological constitutive relation that correlates the shear stress
to the rate of shear strain is given by

\begin{equation}
\tau=\tau_{c}\,\mathrm{asinh}\left(\frac{\mu_{0}\gamma}{\tau_{c}}\right)
\end{equation}
where $\mu_{0}$ is the viscosity at vanishing rate of shear strain, and $\tau_{c}$ is a
characteristic shear stress. Hence, the generalized Newtonian viscosity is given by

\begin{equation}
\mu=\frac{\tau}{\gamma}=\frac{\tau_{c}\,\mathrm{asinh}\left(\frac{\mu_{0}\gamma}{\tau_{c}}\right)}{\gamma}
\end{equation}
On substituting $\mu$ from the last relation into Equation \ref{MainVar} we obtain

\begin{equation}
\frac{d}{d
r}\left(\frac{\tau_{c}\,\mathrm{asinh}\left(\frac{\mu_{0}\gamma}{\tau_{c}}\right)}{\gamma}\frac{d\gamma}{dr}\right)=0
\end{equation}
On integrating once we get

\begin{equation}
\frac{\tau_{c}\,\mathrm{asinh}\left(\frac{\mu_{0}\gamma}{\tau_{c}}\right)}{\gamma}\frac{d\gamma}{dr}=A
\end{equation}
where $A$ is a constant. On separating the two variables and integrating again we obtain

\begin{equation}\label{numInt}
\int\frac{\tau_{c}\,\mathrm{asinh}\left(\frac{\mu_{0}\gamma}{\tau_{c}}\right)}{\gamma}d\gamma=A\int
dr=Ar
\end{equation}
where the constant of integration $C$ is absorbed within the integral on the left hand side. The
left hand side of Equation \ref{numInt} is not a standard integral that can be easily, if possible
at all, to integrate analytically using standard integration methods or can be found in tables of
integrals. On using a computer algebra system, the following expression was obtained

\begin{equation}
\begin{multlined}
\int\frac{\tau_{c}\,\mathrm{asinh}\left(\frac{\mu_{0}\gamma}{\tau_{c}}\right)}{\gamma}d\gamma= \\
\frac{\tau_{c}}{2}\left[\mathrm{asinh^{2}}\left(\frac{\mu_{0}\gamma}{\tau_{c}}\right)+2\,\mathrm{asinh}\left(\frac{\mu_{0}\gamma}{\tau_{c}}\right)\ln\left(1-e^{-2\,\mathrm{asinh}\left(\frac{\mu_{0}\gamma}{\tau_{c}}\right)}\right)-\mathrm{Li}_{2}\left(e^{-2\,\mathrm{asinh}\left(\frac{\mu_{0}\gamma}{\tau_{c}}\right)}\right)\right]\end{multlined}
\end{equation}
where $\mathrm{Li}_{2}$ is the polylogarithm function. This expression, when evaluated, produces
significant errors as it diverges with decreasing $r$ due apparently to accumulated numerical
errors from the logarithmic and polylogarithmic functions.

To solve this problem we used a numerical integration procedure to evaluate this integral, and
hence obtain $A$, numerically using the boundary condition at the tube wall, that is

\begin{equation}
\gamma\left(r=R\right)\equiv\gamma_{w}=\frac{\tau_{c}}{\mu_{0}}\sinh\left(\frac{\tau_{w}}{\tau_{c}}\right)
\end{equation}
where $\tau_{w}$ is the shear stress at the tube wall which is defined as the ratio of the force
normal to the tube cross section, $F_{\perp}$, to the area of the luminal surface parallel to this
force, $A_{\parallel}$, that is

\begin{equation}\label{tauw}
\tau_{w}\equiv\frac{F_{\perp}}{A_{\parallel}}=\frac{\pi R^{2}\Delta p}{2\pi RL}=\frac{R\,\Delta
p}{2L}
\end{equation}
where $R$ and $L$ are the tube radius and length respectively, and $\Delta p$ is the pressure drop
across the tube.

This was then followed by obtaining $\gamma$ as a function of $r$ using a bisection numerical
solver in conjunction with a numerical integration procedure based on Equation \ref{numInt}. Due to
the fact that the constant of integration $C$ is absorbed in the left hand side and a numerical
integration procedure was used rather than an analytical evaluation of the integral on the left
hand side of Equation \ref{numInt}, there was no need for an analytical evaluation of this constant
using the boundary condition at the tube center, i.e.
\begin{equation}
\gamma\left(r=0\right)=0
\end{equation}
The numerically-obtained $\gamma$ was then integrated numerically with respect to $r$ to obtain the
flow velocity as a function of $r$. The flow velocity profile was then integrated numerically with
respect to the cross sectional area to obtain the volumetric flow rate, as outlined previously.

\subsection{Casson Fluids}

For the Casson fluids, the rheological constitutive relation is given by

\begin{equation}
\tau=\left[\left(k\gamma\right)^{1/2}+\tau_{0}^{1/2}\right]^{2}
\end{equation}
where $k$ is a viscosity consistency coefficient, and $\tau_{0}$ is the yield stress. Hence

\begin{equation}
\mu=\frac{\tau}{\gamma}=\frac{\left[\left(k\gamma\right)^{1/2}+\tau_{0}^{1/2}\right]^{2}}{\gamma}
\end{equation}
On substituting $\mu$ from the last equation into the main variational relation, as given by
Equation \ref{MainVar}, we obtain

\begin{equation}
\frac{d}{d
r}\left(\frac{\left[\left(k\gamma\right)^{1/2}+\tau_{0}^{1/2}\right]^{2}}{\gamma}\frac{d\gamma}{dr}\right)=0
\end{equation}
On integrating once we get

\begin{equation}
\frac{\left[\left(k\gamma\right)^{1/2}+\tau_{0}^{1/2}\right]^{2}}{\gamma}\frac{d\gamma}{dr}=A
\end{equation}
where $A$ is a constant, that is

\begin{equation}
\left[k+2\left(\frac{k\tau_{0}}{\gamma}\right)^{1/2}+\frac{\tau_{0}}{\gamma}\right]\frac{d\gamma}{dr}=A
\end{equation}
On separating the two variables and integrating again we obtain

\begin{equation}
\int\left[k+2\left(\frac{k\tau_{0}}{\gamma}\right)^{1/2}+\frac{\tau_{0}}{\gamma}\right]d\gamma=A\int dr
\end{equation}
i.e.

\begin{equation}
k\gamma+4\left(k\tau_{0}\gamma\right)^{1/2}+\tau_{0}\ln\left(\gamma\right)=Ar+C\label{CasMain}
\end{equation}
where $C$ is another constant.

Now, we have two boundary conditions one at the tube center and the other at the tube wall. As for
the first we have

\begin{equation}
\gamma\left(r=0\right)=0
\end{equation}
As the solution should be finite for all values of $r$ including $r=0$, we should have

\begin{equation}
C=0
\end{equation}
As for the second boundary condition we have

\begin{equation}
k\gamma_{w}+4\left(k\tau_{0}\gamma_{w}\right)^{1/2}+\tau_{0}\ln\left(\gamma_{w}\right)=AR
\end{equation}
where the rate of shear strain at the tube wall, $\gamma_w$, is given by

\begin{equation}
\gamma_{w}=\frac{\left[\sqrt{\tau_{w}}-\tau_{0}^{1/2}\right]^{2}}{k}
\end{equation}
with the shear stress at the tube wall, $\tau_w$, being given by Equation \ref{tauw}. Hence

\begin{equation}
A=\frac{k\gamma_{w}+4\left(k\tau_{0}\gamma_{w}\right)^{1/2}+\tau_{0}\ln\left(\gamma_{w}\right)}{R}
\end{equation}
Equation \ref{CasMain}, which defines $\gamma$ implicitly in terms of $r$, is then used in
conjunction with a numerical bisection method to find $\gamma$ as a function of $r$. The
numerically found $\gamma$ is then integrated numerically with respect to $r$ to obtain the flow
velocity profile which is then integrated numerically with respect to the cross sectional area to
find the volumetric flow rate.

\section{Results and Analysis}

\begin{figure}
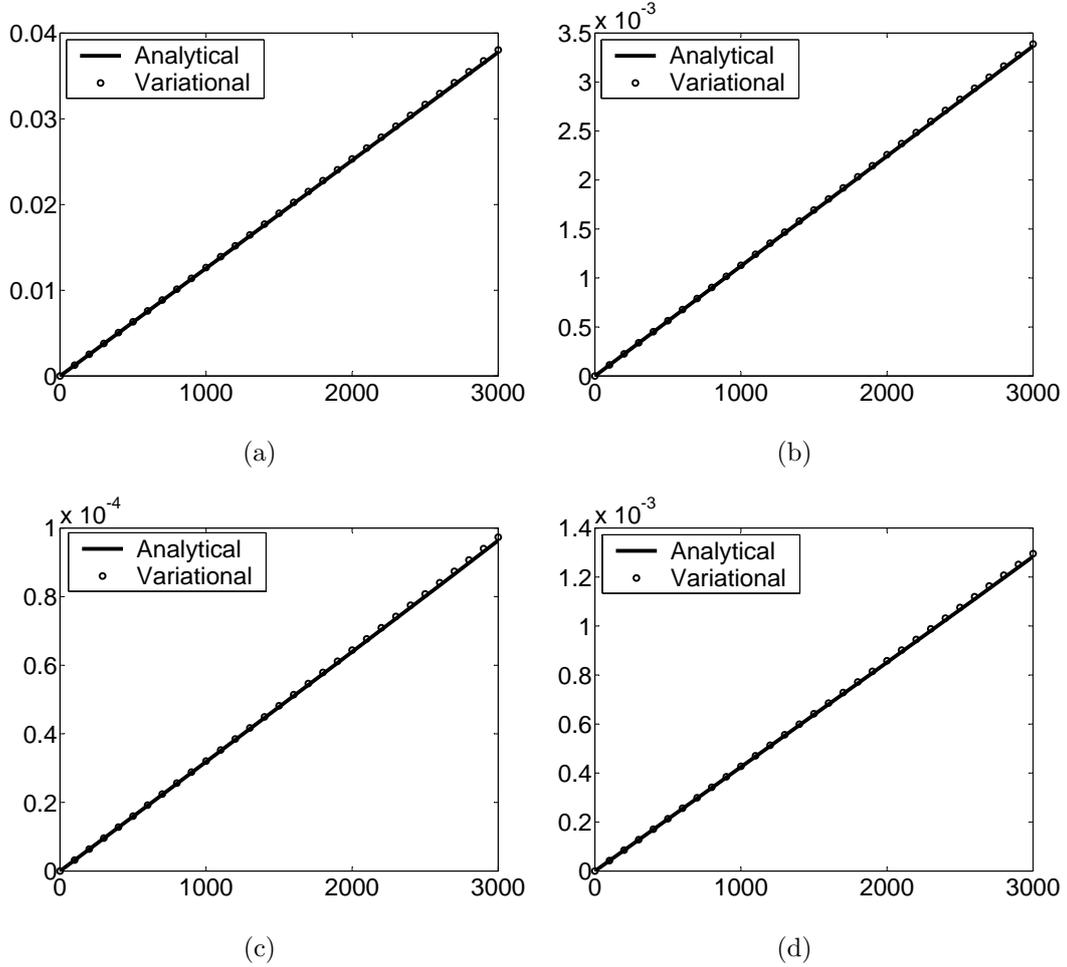

\centering %
\subfigure[]%
{\begin{minipage}[b]{0.5\textwidth} \CIF {g/ReeEyringFig1}
\end{minipage}}
\Hs
\subfigure[]%
{\begin{minipage}[b]{0.5\textwidth} \CIF {g/ReeEyringFig2}
\end{minipage}} \Vmin
\centering %
\subfigure[]%
{\begin{minipage}[b]{0.5\textwidth} \CIF {g/ReeEyringFig3}
\end{minipage}}
\Hs
\subfigure[]%
{\begin{minipage}[b]{0.5\textwidth} \CIF {g/ReeEyringFig4}
\end{minipage}}
\caption{Comparing the WRM analytical solution, as given by Equation \ref{REAnalEq}, to the
variational solution for the flow of Ree-Eyring fluids in circular straight uniform rigid tubes
with
(a) $R=0.02$~m, $L=0.5$~m, $\mu_0=0.01$~Pa.s, and $\tau_c=500$~Pa,
(b) $R=0.01$~m, $L=0.7$~m, $\mu_0=0.005$~Pa.s, and $\tau_c=750$~Pa,
(c) $R=0.003$~m, $L=0.05$~m, $\mu_0=0.02$~Pa.s, and $\tau_c=300$~Pa,
and (d) $R=0.006$~m, $L=0.08$~m, $\mu_0=0.015$~Pa.s, and $\tau_c=400$~Pa.
In all four sub-figures, the vertical axis represents the volumetric flow rate, $Q$, in
m$^3$.s$^{-1}$ while the horizontal axis represents the pressure drop, $\Delta p$, in Pa. The
average percentage relative difference between the variational and the WRM analytical solutions for
these cases are about 0.67\%, 0.65\%, 0.77\%, and 0.75\% respectively. \label{ReeEyringFig}}
\end{figure}

\begin{figure}
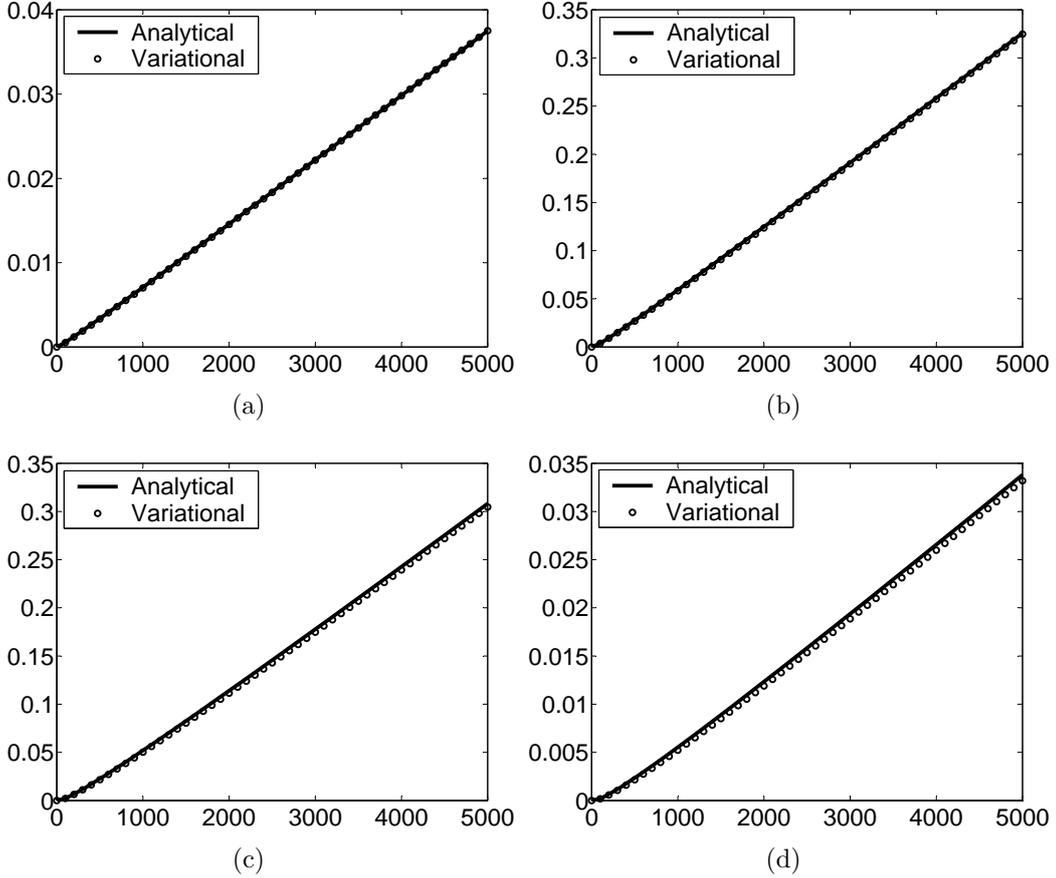

\centering %
\subfigure[]%
{\begin{minipage}[b]{0.5\textwidth} \CIF {g/CassonFig1}
\end{minipage}}
\Hs
\subfigure[]%
{\begin{minipage}[b]{0.5\textwidth} \CIF {g/CassonFig2}
\end{minipage}} \Vmin
\centering %
\subfigure[]%
{\begin{minipage}[b]{0.5\textwidth} \CIF {g/CassonFig3}
\end{minipage}}
\Hs
\subfigure[]%
{\begin{minipage}[b]{0.5\textwidth} \CIF {g/CassonFig4}
\end{minipage}}
\caption{Comparing the WRM analytical solution, as given by Equation \ref{CasAnalEq}, to the
variational solution for the flow of Casson fluids in circular straight uniform rigid tubes with
(a) $R=0.01$~m, $L=0.1$~m, $k=0.005$~Pa.s, and $\tau_o=0.1$~Pa,
(b) $R=0.05$~m, $L=0.5$~m, $k=0.07$~Pa.s, and $\tau_o=0.25$~Pa,
(c) $R=0.05$~m, $L=0.5$~m, $k=0.07$~Pa.s, and $\tau_o=0.75$~Pa,
and (d) $R=0.01$~m, $L=0.1$~m, $k=0.005$~Pa.s, and $\tau_o=1.0$~Pa.
In all four sub-figures, the vertical axis represents the volumetric flow rate, $Q$, in
m$^3$.s$^{-1}$ while the horizontal axis represents the pressure drop, $\Delta p$, in Pa. The
average percentage relative difference between the variational and the WRM analytical solutions for
these cases are 0.55\%, 1.03\%, 2.26\%, and 4.07\% respectively. \label{CassonFig}}
\end{figure}

The method, as described in the last section, was implemented and applied to the Ree-Eyring and
Casson fluids over an extended range of the tube and fluid parameters, and the results were
examined and analyzed. In all cases we obtained very good agreements between the variational
solutions and the WRM analytical solutions; a sample of the investigated cases are presented in
Figures \ref{ReeEyringFig} and \ref{CassonFig}.

The difference between the two solutions for the Ree-Eyring fluids can be largely explained by
numerical errors arising from the extensive use of numerical integration and bisection solvers in
several stages of the method implementation. These errors propagate and amplify on passing from one
stage to the next. However, almost in all cases that have been investigated, the maximum and the
average of the percentage relative difference between the variational and the WRM analytical
solutions do not exceed 1\%.

As for the Casson fluids, as well as the above mentioned accumulated and magnified errors from the
persistent use of numerical integration and numerical solvers in multiple sequential stages, there
is a more fundamental reason for the deviation between the two solutions, that is the variational
method is strictly applicable to real fluids with no yield stress and hence the use of this method
for the viscoplastic materials is just an approximation which is usually a good one when the value
of the yield stress is low. Hence, in Figure \ref{CassonFig} we can see an obvious and logical
trend that is as the yield stress value increases, the deviation between the variational and the
WRM analytical solutions increases as it should be because the approximation in using the
variational method is worsened as the value of the yield stress rises. As the yield stress
increases, the non-yielded plug region in the middle of the tube, which is not strictly subject to
the variational formulation, increases in size and hence the gap between the two solutions is
widened.


\clearpage
\section{Conclusions} \label{Conclusions}

In this paper, we demonstrated the use of the variational method that optimizes the total stress of
the fluid in the flow conduit to obtain flow relations correlating the volumetric flow rate to the
pressure drop for the flow of two non-Newtonian fluid models, namely Ree-Eyring and Casson, in
rigid straight uniform tubes with circular cross sectional shape. The variational method was
closely examined and analyzed using extensive ranges of fluid and conduit parameters and the
results were compared to the analytically-obtained solutions from the
Weissenberg-Rabinowitsch-Mooney equation. In all cases, good agreement was observed between the
variational and the WRM analytical solutions within anticipated and justifiable numerical errors
and modeling approximations.

Although the variational principle in its current formulation strictly applies only to
non-viscoplastic fluids, the analysis shows that the variational method is a good approximation for
the viscoplastic fluids with low value of yield stress as demonstrated by several examples from the
Casson fluids. An obvious logical trend was observed where the gap between the variational and the
WRM analytical solutions was enlarged as the value of the yield stress of the viscoplastic fluids
was increased.

Overall, this investigation adds more support to the previous investigations
\cite{SochiVariational2013, SochiSlitPaper2014} related to the application of the Euler-Lagrange
variational principle to the flow of generalized Newtonian fluids through rigid straight uniform
pipes with circular cross sections and slits of a plane long thin geometry. Further investigations
are planned for the future to extend the application of this method to other types of fluid, flow
condition and flow conduit to verify the validity of the variational principle and the
applicability of the derived method to the flow dynamic systems in general. As part of this
investigation, the volumetric flow rate relations for the Ree-Eyring and Casson fluids in circular
tubes were derived using the well-known Weissenberg-Rabinowitsch-Mooney equation.

\vspace{0.5cm}
\phantomsection \addcontentsline{toc}{section}{Nomenclature} %
{\noindent \LARGE \bf Nomenclature} \vspace{0.5cm}

\begin{supertabular}{ll}
$\gamma$                &   rate of shear strain \\
$\gamma_w$              &   rate of shear strain at tube wall \\
$\mu$                   &   viscosity of generalized Newtonian fluids \\
$\mu_0$                 &   viscosity at zero shear rate in Ree-Eyring model \\
$\tau$                  &   shear stress \\
$\tau_0$                &   yield stress in Casson model \\
$\tau_c$                &   characteristic shear stress in Ree-Eyring model \\
$\tau_m$                &   shear stress at tube center \\
$\tau_{t}$              &   total shear stress \\
$\tau_w$                &   shear stress at tube wall \\
\\
$A_{\parallel}$         &   area of tube luminal surface \\
$F_{\perp}$             &   force normal to tube cross section \\
$k$                     &   viscosity consistency coefficient in Casson model \\
$L$                     &   tube length \\
Li$_2$                  &   polylogarithm function \\
$p$                     &   pressure \\
$\Delta p$              &   pressure drop \\
$Q$                     &   volumetric flow rate \\
$r$                     &   radius \\
$R$                     &   tube radius \\

\end{supertabular}

\newpage
\phantomsection \addcontentsline{toc}{section}{References} %
\bibliographystyle{unsrt}

\appendix

\clearpage
\section{Deriving Flow Relations for Ree-Eyring and Casson Fluids}\label{AppA}

In this appendix we derive analytical expressions correlating the volumetric flow rate to the
pressure drop for the flow of Ree-Eyring and Casson fluids in rigid straight uniform tubes with
circular cross sections which we could not find in the literature. For this purpose we use the
Weissenberg-Rabinowitsch-Mooney relation which is given by \cite{Skellandbook1967}

\begin{equation}
Q=\frac{\pi R^{3}}{\tau_{w}^{3}}\int_{0}^{\tau_{w}}\tau^{2}\gamma d\tau
\end{equation}
where $Q$ is the volumetric flow rate, $\tau$ is the shear stress, $\gamma$ is the rate of shear
strain, and $\tau_{w}$ is the shear stress at the tube wall which is given by

\begin{equation}
    \tau_w = \frac {R \, \Delta p} {2 L}
\end{equation}
where $R$ and $L$ are the tube radius and length respectively, and $\Delta p$ is the pressure drop
across the tube.

\vspace{0.7cm}

For the Ree-Eyring fluids we have

\begin{equation}
\tau=\tau_{c} \, \mathrm{asinh}\left(\frac{\mu_{0}\gamma}{\tau_{c}}\right)
\end{equation}
where $\mu_{0}$ is the viscosity at vanishing rate of shear strain, and $\tau_{c}$ is a
characteristic shear stress. Therefore

\begin{equation}
\gamma=\frac{\tau_{c}}{\mu_{0}}\sinh\left(\frac{\tau}{\tau_{c}}\right)
\end{equation}
Inserting this into the Weissenberg-Rabinowitsch-Mooney relation we obtain

\begin{equation}
Q=\frac{\pi R^{3}\tau_{c}}{\tau_{w}^{3}\mu_{0}}\int_{0}^{\tau_{w}}\tau^{2}\sinh\left(\frac{\tau}{\tau_{c}}\right)d\tau
\end{equation}
that is

\begin{equation}\label{REAnalEq}
\boxed{
Q=\frac{\pi
R^{3}\tau_{c}}{\tau_{w}^{3}\mu_{0}}\left[\left(\tau_{c}\tau_{w}^{2}+2\tau_{c}^{3}\right)\cosh\left(\frac{\tau_{w}}{\tau_{c}}\right)-2\tau_{c}^{2}\tau_{w}\sinh\left(\frac{\tau_{w}}{\tau_{c}}\right)-2\tau_{c}^{3}\right]
}
\end{equation}

\vspace{0.7cm}

For the Casson fluids we have

\begin{equation}
\tau^{1/2}=\left(k\gamma\right)^{1/2}+\tau_{0}^{1/2}
\end{equation}
where $k$ is a viscosity consistency coefficient, and $\tau_{0}$ is the yield stress. Hence

\begin{equation}
\gamma=\frac{\left(\tau^{1/2}-\tau_{0}^{1/2}\right)^{2}}{k}
\end{equation}
On substituting this into the Weissenberg-Rabinowitsch-Mooney relation we obtain

\begin{equation}
Q=\frac{\pi R^{3}}{\tau_{w}^{3}k}\int_{0}^{\tau_{w}}\tau^{2}\left(\tau^{1/2}-\tau_{0}^{1/2}\right)^{2}d\tau
\end{equation}
that is

\begin{equation}\label{CasAnalEq}
\boxed{
Q=\frac{\pi
R^{3}}{\tau_{w}^{3}k}\left(\frac{\tau_{w}^{4}}{4}-\frac{4\sqrt{\tau_{0}}\tau_{w}^{7/2}}{7}+\frac{\tau_{0}\tau_{w}^{3}}{3}\right)
}
\end{equation}

\end{document}